\documentclass[conference]{IEEEtran}

\usepackage[left=1.6cm,right=1.6cm,top=1.8cm,bottom=1.8cm]{geometry}

\usepackage{algorithm}	
\usepackage{algpseudocode}

\usepackage[nospace]{cite}	

\usepackage[caption=false,font=footnotesize]{subfig}	
\usepackage{graphicx}	

\usepackage{amsmath}
\usepackage{amsfonts}
\usepackage{amssymb}
\usepackage{amsthm}
\usepackage{mathtools}

\usepackage{array}		
\usepackage{subdepth}	

\usepackage{url}	

\usepackage{xspace}	
\usepackage{xcolor}	

\usepackage{tikz}
\usetikzlibrary{arrows,decorations.pathreplacing}

\usepackage[straightvoltages]{circuitikz}

\newcommand{\tea}{t}

\newcommand{\Sec}[1][]{Sec#1.}
\newcommand{\Fig}[1][]{Fig#1.}
\newcommand{\Tab}[1][]{Tab#1.}
\newcommand{\Alg}[1][]{Alg#1.}
\newcommand{\Hyp}[1][]{Hyp#1.}

\newcommand{\estim}[1]{\widehat{#1}}
\newcommand{\meas}[1]{\widetilde{#1}}

\newcommand{\Set}[1]{\mathcal{#1}}
\newcommand{\Card}[1]{\left|#1\right|}

{
\theoremstyle{plain}
\newtheorem{hypothesis}{Hypothesis}

\newtheorem{lemma}{Lemma}
}

\newcommand{\diff}{\operatorname{D}}

\newcommand{\sign}{\operatorname{sign}}

\newcommand{\Conj}[1]{#1^{*}}
\newcommand{\Real}[1]{\Re\left\{#1\right\}}
\newcommand{\Imag}[1]{\Im\left\{#1\right\}}
\newcommand{\Abs}[1]{\left|#1\right|}

\newcommand{\Compl}[1]{#1_{\complement}}

\newcommand{\cartprod}{\times}		
\newcommand{\hadprod}{\circ}		
\newcommand{\schurcomp}{/}			

\newcommand{\col}{\operatorname{col}}
\newcommand{\diag}{\operatorname{diag}}

\newcommand{\Matrix}[1]{\boldsymbol{\mathbf{#1}}}
\newcommand{\Tran}[1]{#1^{\top}}
\newcommand{\Inv}[1]{#1^{-1}}
\newcommand{\Norm}[1]{\left\lVert#1\right\rVert}

\newcommand{\Graph}[1]{\mathfrak{#1}}

\newcommand{\IEEE}{IEEE\xspace}			
\newcommand{\EPFL}{EPFL\xspace}	

\newcommand{\MANA}{MANA\xspace}			
\newcommand{\EMTP}[1][]{EMTP#1\xspace}	

\newcommand{\PMU}[1][]{PMU#1\xspace}	
\newcommand{\FSR}[1]{FSR#1\xspace}		
\newcommand{\RMS}{RMS\xspace}				

\newcommand{\PFS}{PFS\xspace}				
\newcommand{\PFE}[1][]{PFE#1\xspace}		
\newcommand{\GSM}[1][]{GSM#1\xspace}	
\newcommand{\NRM}[1][]{NRM#1\xspace}	

\newcommand{\SE}{SE\xspace}			
\newcommand{\LSR}{LSR\xspace}		
\newcommand{\WLSR}{W\LSR}			
\newcommand{\KF}[1][]{KF\xspace}		

\newcommand{\VSA}{VSA\xspace}			
\newcommand{\CM}[1][]{CM#1\xspace}	
\newcommand{\CPF}{CPF\xspace}			
\newcommand{\VSI}[1][]{VSI#1\xspace}	

\newcommand{\KR}{KR\xspace}				
\newcommand{\TE}[1][]{TE#1\xspace}		
\newcommand{\PM}[1][]{PM#1\xspace}	

\begin{document}


\title{Performance Assessment of Kron Reduction in the Numerical Analysis of Polyphase Power Systems}


%

\author
{%
	\IEEEauthorblockN{Andreas~Martin~Kettner and Mario~Paolone}%
	\IEEEauthorblockA%
	{%
		{\'E}cole~Polytechnique~F{\'e}d{\'e}rale~de~Lausanne (\EPFL)\\
		Lausanne,~Vaud,~Switzerland%
	}
}


\newcommand{\I}{\Matrix{I}}
\newcommand{\Y}{\Matrix{Y}}
\newcommand{\V}{\Matrix{V}}

\maketitle



\begin{abstract}
	This paper investigates the impact of Kron reduction on the performance of numerical methods applied to the analysis of unbalanced polyphase power systems.
	Specifically, this paper focuses on power-flow study, state estimation, and voltage stability assessment.
	For these applications, the standard Newton-Raphson method, linear weighted-least-squares regression, and homotopy continuation method are used, respectively.
	The performance of the said numerical methods is assessed in a series of simulations, in which the zero-injection nodes of a test system are successively eliminated through Kron reduction.
\end{abstract}






\section{Introduction}

Any application in power system analysis, including \emph{Power-Flow Study} (\PFS), \emph{State Estimation} (\SE), and \emph{Voltage Stability Assessment} (\VSA), inherently relies on equivalent circuits of the power system components.
Usually, power grids consist of exactly linear components (e.g., lines) or approximately linear components (e.g., transformers), which can be represented by linear equivalent circuits (e.g., by admittance parameters).
Other components, such as generators or loads, do have to be represented by nonlinear equivalent circuits (e.g., to account for power control).
Thus, the power system model is described by a system of both linear and nonlinear equations, which can be solved using numerical methods.


Evidently, the computational burden of the aforementioned numerical methods scales with the number of unknowns (i.e., phasors of nodal voltages and/or branch currents).
Therefore, model reduction techniques are often employed.
In particular, \emph{Kron reduction} (\KR) is commonly used \cite{B:PSE:CA:1959:Kron,J:PSE:CA:2013:Doerfler}.
Fundamentally, \KR eliminates nodes with zero current injection from the grid, thereby reducing the number of linear equations (i.e., the order of the admittance matrix).
Recently, the authors of this paper performed a rigorous analysis of the feasibility of \KR for monophase and polyphase power grids \cite{J:PSE:CA:2018:Kettner:1,J:PSE:CA:2018:Kettner:2}.
In line with the said analysis, this paper assesses the impact of \KR on the performance of state-of-the-art numerical methods applied to the analysis of polyphase power systems.
To be more precise, \PFS, \SE, and \VSA are considered.



The rest of the paper is organized as follows.
First, a review of the literature is presented (\Sec\,\ref{sec:lit}), and the system model is described (\Sec\,\ref{sec:mod}).
Then, the utilized numerical methods are discussed (\Sec\,\ref{sec:meth}), and their performance is assessed (\Sec\,\ref{sec:perf}).
Finally, the conclusions are drawn (\Sec\,\ref{sec:concl}).

\section{Literature Review}
\label{sec:lit}


\subsection{Power-Flow Study}
\label{sec:lit:pfs}


In \PFS, the use of positive-sequence equivalent circuits is common \cite{J:PSE:CA:1918:Fortescue}.
An overview of numerical methods for solving the \emph{Power-Flow Equations} (\PFE[s]) is provided in \cite{J:PSE:PFS:1974:Stott}. 
Notably, fixed-point techniques, like the \emph{Gauss-Seidel} and \emph{Newton-Raphson method} (\GSM/\NRM) \cite{J:PSE:PFS:1970:Meisel}, or the conjugate-gradient and steepest-descent method \cite{J:PSE:PFS:1968:Wallach}, are popular.


Naturally, \PFS can also be performed for equivalent circuits of polyphase power systems \cite{J:PSE:CA:1968:Laughton}.
Indeed, the treatment of the polyphase case is similar to the monophase case.
For instance, the \PFE[s] can be written in fixed-point form \cite{J:PSE:PFS:2018:Wang}, and be solved using the \GSM \cite{J:PSE:PFS:1982:Tiwari} or \NRM \cite{J:PSE:PFS:1974:Wasley}.
Another method \cite{J:PSE:PFS:2014:Kocar}, which is notably implemented in \EMTP[-RV], is based on the so-called modified augmented nodal analysis (a.k.a. \MANA).


\subsection{State Estimation}
\label{sec:lit:se}

Here, the estimation of the \emph{steady state} (i.e., phasors of nodal voltages and/or branch currents) is considered.

%
\SE typically uses positive-sequence equivalent circuits, too.
A survey of methods is presented in \cite{J:PSE:SE:2000:Monticelli}.
Notably, nonlinear or linear \emph{Weighted-Least-Squares Regression} (\WLSR) \cite{J:PSE:SE:1970:Larson,J:PSE:SE:1970:Schweppe:1,J:PSE:SE:1970:Schweppe:2} and the \emph{Kalman Filter} (\KF) \cite{J:PSE:SE:1970:Debs} are popular.
%
%
These methods have also been applied polyphase power systems \cite{B:PSE:SE:2016:Paolone,J:PSE:SE:2017:Kettner}.


\subsection{Voltage Stability Assessment}
\label{sec:lit:vsa}

The term \emph{voltage stability} refers to a variety of phenomena, ranging from the transient to steady-state timescale \cite{J:PSE:VSA:2004:TF}.
Here, \emph{steady-state voltage stability}, which is related to the solvability of the \PFE[s], is considered.


Classical \VSA also works with positive-sequence equivalent circuits.
\emph{Continuation Methods} (\CM[s]) (e.g., \cite{J:PSE:VSA:1992:Ajjarapu,J:PSE:VSA:1993:Canizares,J:PSE:VSA:1995:Chiang}), a.k.a. \emph{Continuation Power Flow} (\CPF), determine a critical operating point by producing a continuum of \PFE solutions.
The stability margin can be obtained by solving a nonlinear program \cite{J:PSE:VSA:1997:Irisarri}.
Alternatively, the solvability of the \PFE[s] can be assessed using \emph{Voltage Stability Indices} (\VSI[s]), like the singular values \cite{J:PSE:VSA:1993:Loef}, eigenvalues \cite{J:PSE:VSA:1992:Gao}, or determinant \cite{J:PSE:VSA:1998:Prada} of the Jacobian matrix.
%
%
Notably, \CPF methods \cite{J:PSE:VSA:2014:Sheng} and \VSI[s] \cite{J:PSE:VSA:2018:Kettner} have been applied to polyphase power systems.


\subsection{Contribution of this Paper}
\label{sec:lit:contrib}

This paper investigates the impact of \KR on the performance of state-of-the-art numerical methods for \PFS, \SE, and \VSA, which are applied to the analysis of unbalanced polyphase power systems.
More precisely, the standard \NRM is used for \PFS, linear \WLSR for \SE, and the homotopy \CM for \VSA.
\clearpage

\section{System Model}
\label{sec:mod}

This section is a summary of the models discussed in \cite{J:PSE:CA:2018:Kettner:2,J:PSE:VSA:2018:Kettner}.
Unless stated otherwise, quantities are expressed in per unit.


\subsection{Electrical Grid}
\label{sec:mod:grid}

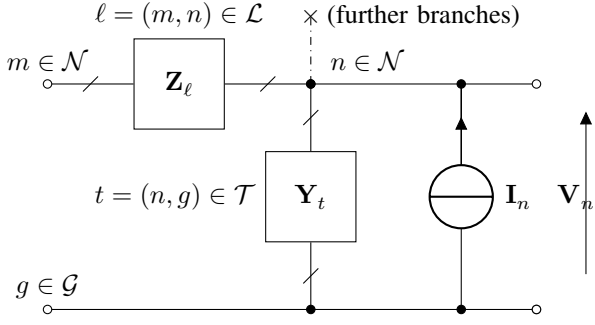
\begin{figure}[t]
	\centering
	\tikzstyle{block}=[rectangle, draw=black,minimum size=1.2cm]

\begin{circuitikz}
	
	\def\TerminalHorizontal{3.5}	
	\def\SourceHorizontal{2.0}
	\def\PhaseVertical{3.0}
	\def\LabelSeparation{0.3}
	\def\TickSize{0.1}
	
	
	\coordinate (PL) at (-\TerminalHorizontal,\PhaseVertical);
	\coordinate (PC) at (0,\PhaseVertical);
	\coordinate (PR) at ($(1.5*\SourceHorizontal,\PhaseVertical)$);
	
	\coordinate (NL) at (-\TerminalHorizontal,0);	
	\coordinate (NC) at (0,0);
	\coordinate (NR) at ($(1.5*\SourceHorizontal,0)$);
	
	
	
	\draw (NL) to[short,o-o] (NR);
	\node at ($(NL)+\LabelSeparation*(0,1)$) {$g\in\Set{G}$};
	

	
	\node[block] (ZL) at (-0.5*\TerminalHorizontal,\PhaseVertical) {$\Matrix{Z}_{\ell}$};
	
	\draw (PL)
		to[short,o-] (ZL.west)
		to[open] (ZL.east)
		to[short,-*] (PC);
	
	\node at ($(PL)+\LabelSeparation*(0,1)$) {$m\in\mathcal{N}$};
	\node at ($(ZL.north)+\LabelSeparation*(0,1)$) {$\ell=(m,n)\in\mathcal{L}$};
	
	\draw ($0.5*(PL)+0.5*(ZL.west)-\TickSize*(1,1)$) to +($2*\TickSize*(1,1)$);
	\draw ($0.5*(PC)+0.5*(ZL.east)-\TickSize*(1,1)$) to +($2*\TickSize*(1,1)$);
		
	
	
	\node[block] (YT) at (0,0.5*\PhaseVertical) {$\mathbf{Y}_{\tea}$};
	\coordinate (UC) at ($(PC)+0.3*\PhaseVertical*(0,1)$);
	
	\draw (PC)
		to[short,*-] (YT.north)
		to[open] (YT.south)
		to[short,-*] (NC);
	\draw[dashdotted] (PC) to (UC);
	
	\node at ($(PC)+\LabelSeparation*(2.5,1)$) {$n\in\mathcal{N}$};
	\node at ($(YT.west)-\LabelSeparation*(4,0)$) {$\tea=(n,g)\in\mathcal{T}$};
	\node at (UC) {$\times$};
	\node at ($(UC)+\LabelSeparation*(5,0)$) {(further branches)};
	
	\draw ($0.5*(PC)+0.5*(YT.north)-\TickSize*(1,1)$) to +($2*\TickSize*(1,1)$);
	\draw ($0.5*(NC)+0.5*(YT.south)-\TickSize*(1,1)$) to +($2*\TickSize*(1,1)$);
	
	
	
	\draw (PR) to (PC);
	\draw (NR) to[open,v=$\Matrix{V}_{n}$,o-o] (PR);
	\draw (\SourceHorizontal,0) to[current source,i_=$\Matrix{I}_{n}$,*-*] (\SourceHorizontal,\PhaseVertical);
	
\end{circuitikz}
	\caption
	{%
		Definition of the compound branch impedance matrices $\Matrix{Z}_{\ell}$ ($\ell\in\Set{L}$), compound shunt admittance matrices $\Matrix{Y}_{\tea}$ ($\tea\in\Set{T}$), nodal phase-to-ground voltage vectors $\Matrix{V}_{n}$, and injected current vectors $\Matrix{I}_{n}$ ($n\in\Set{N}$).
	}
	\label{fig:mod:grid}
\end{figure}


Consider a generic polyphase power grid, which is equipped with a neutral conductor.
The system is wired as follows:
\begin{hypothesis}
	\label{hyp:neutral}
	The neutral conductor is effectively grounded (i.e., the neutral-to-ground voltage is zero),
	and it is connected to the reference points of all voltage or current sources.
\end{hypothesis}
\noindent
That is, the phase-to-neutral voltages are effectively referenced w.r.t. the ground, and fully describe the system.



The ground node is denoted by $\Set{G}$, and the polyphase nodes, each of which has a complete set of phase terminals $\Set{P}$, by $\Set{N}$.
The equivalent circuit of the grid is composed of polyphase branches $\Set{L}\subseteq\Set{N}\times\Set{N}$ and polyphase shunts $\Set{T}=\Set{N}\times\Set{G}$, which are characterized by compound branch impedance matrices $\Matrix{Z}_{\ell}$ ($\ell\in\Set{L}$) and compound shunt admittance matrices $\Matrix{Y}_{\tea}$ ($\tea\in\Set{T}$), respectively (see \Fig~\ref{fig:mod:grid}).
It is supposed that
\begin{hypothesis}
	\label{hyp:params}
	The compound electrical parameters satisfy
	\begin{alignat}{4}
		\forall \Matrix{Z}_{\ell}
		&:				~&
		\Matrix{Z}_{\ell}
		&=			\Tran{\Matrix{Z}_{\ell}},~&
		\exists\Matrix{Y}_{\ell}
		&=			\Inv{\Matrix{Z}_{\ell}},~&
		\Real{\Matrix{Z}_{\ell}}
		&\succeq	0
		\label{eq:params:branch}
		\\
		\text{if}~\Matrix{Y}_{\tea}\neq\Matrix{0}
		&:				~&
		\Matrix{Y}_{\tea}
		&=			\Tran{\Matrix{Y}_{\tea}},~&
		\exists\Matrix{Z}_{\tea}
		&=			\Inv{\Matrix{Y}_{\tea}},~&
		\Real{\Matrix{Y}_{\tea}}
		&\succeq	0
		\label{eq:params:shunt}
	\end{alignat}
\end{hypothesis}
\noindent
The topology is described by the branch graph $\Graph{B}\coloneqq(\Set{N},\Set{L})$ and the shunt graph $\Graph{S}\coloneqq(\Set{N}\cup\Set{G},\Set{T})$.



Let $V_{n,p}$ and $I_{n,p}$ the phasors of the phase-to-ground voltage and injected current in phase $p$ of node $n$, respectively.
Define corresponding vectors for the nodes (see \Fig~\ref{fig:mod:grid}) and the grid
\begin{alignat}{2}
	\Matrix{V}_{n}
	&\coloneqq	\col_{p\in\Set{P}}\left(V_{n,p}\right),~&
	\Matrix{V}
	&\coloneqq	\col_{n\in\Set{N}}\left(\Matrix{V}_{n}\right)
	\label{eq:grid:volt}
	\\
	\Matrix{I}_{n}
	&\coloneqq	\col_{p\in\Set{P}}\left(I_{n,p}\right),~&
	\Matrix{I}
	&\coloneqq	\col_{n\in\Set{N}}\left(\Matrix{I}_{n}\right)
	\label{eq:grid:cur}
\end{alignat}
Further, let $\Matrix{A}_{\Graph{B}}$ be the (edge-to-vertex) incidence matrix of $\Graph{B}$.
As this matrix is uniquely defined irrespective of the topology, the proposed model applies to both radial and meshed grids.
The primitive compound admittance matrices $\Matrix{Y}_{\Set{L}}$ and $\Matrix{Y}_{\Set{T}}$, and the polyphase incidence matrix $\Matrix{A}_{\Graph{B}}^{\Set{P}}$ are defined as
\begin{align}
	\Matrix{Y}_{\Set{L}}
	&\coloneqq	\diag_{\ell\in\Set{L}}\left(\Matrix{Y}_{\ell}\right),~
	\Matrix{Y}_{\Set{T}}
	\coloneqq		\diag_{\tea\in\Set{T}}\left(\Matrix{Y}_{\tea}\right)
	\\
	\Matrix{A}_{\Graph{B}}^{\Set{P}}
	&\coloneqq	\Matrix{A}_{\Graph{B}} \otimes \diag\left(\Matrix{1}_{\Card{\Set{P}}}\right) 
\end{align}
where $\otimes$ is the Kronecker product, and $\Matrix{1}_{\Card{\Set{P}}}$ is a column vector of ones with length $\Card{\Set{P}}$.
The compound admittance matrix $\Matrix{Y}$, which establishes the relation $\Matrix{I}=\Matrix{Y}\Matrix{V}$, is given by
\begin{equation}
	\Matrix{Y}
	=		\Tran{(\Matrix{A}_{\Graph{B}}^{\Set{P}})}\Matrix{Y}_{\Set{L}}\Matrix{A}_{\Graph{B}}^{\Set{P}}
		+	\Matrix{Y}_{\Set{T}}
\end{equation}
Accordingly, the injected powers $\Matrix{S}(\Matrix{V})$ are given by
\begin{equation}
	\Matrix{S}(\Matrix{V})	\coloneqq	\Matrix{V} \hadprod \Conj{(\Matrix{Y}\Matrix{V})}
	\label{eq:grid:pwr}
\end{equation}
where $\hadprod$ is the Hadamard product and $\Conj{}$ the complex conjugate.


Let $\Set{A}$ and $\Set{B}$ be nonempty disjoint subsets of $\Set{N}$.
Moreover, let $\Matrix{I}_{\Set{A}}$, $\Matrix{V}_{\Set{B}}$, and $\Matrix{Y}_{\Set{A}\cartprod\Set{B}}$ be the associated blocks of $\Matrix{I}$, $\Matrix{V}$, and $\Matrix{Y}$, respectively.
It holds that (for proof, see \cite{J:PSE:CA:2018:Kettner:2}):
\begin{lemma}[Kron Reduction]
	\label{Thm:Kron}
	Let $\Set{Z}\subsetneq\Set{N},~\Set{Z}\neq\emptyset$ s.t. $\Matrix{I}_{\Set{Z}}=\Matrix{0}$, and $\Set{Z}_{\complement}\coloneqq\Set{N}\setminus\Set{Z}$.
	If \Hyp[s]~\ref{hyp:neutral}--\ref{hyp:params} hold, $\Graph{B}$ is weakly connected, and $\Re\{\Matrix{Z}_{\ell}\}\succ0$ $\forall\ell\in\Set{L}$, then
	$\Matrix{V}_{\Set{Z}}$ is a linear function of $\Matrix{V}_{\Compl{\Set{Z}}}$
	\begin{equation}
		\Matrix{V}_{\Set{Z}}
		=	-\Inv{\Matrix{Y}}_{\Set{Z}\times\Set{Z}}\Matrix{Y}_{\Set{Z}\times\Compl{\Set{Z}}}\Matrix{V}_{\Compl{\Set{Z}}}
		\label{eq:invkron}
	\end{equation}
	As a result, $\Matrix{I}=\Matrix{Y}\Matrix{V}$ can be reduced to
	\begin{equation}
		\Matrix{I}_{\Compl{\Set{Z}}} 
		=	(\Matrix{Y} \schurcomp \Matrix{Y}_{\Set{Z}\times\Set{Z}}) \Matrix{V}_{\Compl{\Set{Z}}}
		\label{eq:kron}
	\end{equation}
	where $\Matrix{Y} / \Matrix{Y}_{\Set{Z}\times\Set{Z}}$ is the Schur complement of $\Matrix{Y}$ w.r.t. $\Matrix{Y}_{\Set{Z}\times\Set{Z}}$.
\end{lemma}


\subsection{Aggregate Node Behavior}
\label{sec:mod:node}

The nodes $\mathcal{N}$ are divided into slack nodes $\mathcal{S}$, resource nodes $\mathcal{R}$, and zero-injection nodes $\mathcal{Z}$ (i.e., $\Set{N}=\Set{S}\cup\Set{R}\cup\Set{Z}$).


Slack nodes are represented by \emph{Th{\'e}venin Equivalents} (\TE[s]).
The \TE of slack node $s\in\Set{S}$ consists of a voltage source $\Matrix{V}_{\text{\TE},s}$ and an impedance $\Matrix{Z}_{\text{\TE},s}$.
Supposing $\exists\Matrix{Y}_{\text{\TE},s}\coloneqq\Inv{\Matrix{Z}_{\text{\TE},s}}$, define
\begin{align}
	\Matrix{S}_{\text{\TE},s}(\Matrix{V}_{s})
	&\coloneqq	\Matrix{V}_{s}\hadprod\Conj{\left(\Matrix{Y}_{\text{\TE},s}(\Matrix{V}_{\text{\TE},s}-\Matrix{V}_{s})\right)}
	\\
	\Matrix{S}_{\text{\TE}}(\Matrix{V}_{\Set{S}})
	&\coloneqq	\col_{s\in\Set{S}}(\Matrix{S}_{\text{\TE},s}(\Matrix{V}_{s}))
	\label{eq:te:pwr}
\end{align}


Resource nodes are modeled by \emph{Polynomial Models} (\PM[s]).
The power injected into phase $p\in\Set{P}$ of resource node $r\in\Set{R}$ is represented by a quadratic polynomial
\begin{align}
	\Matrix{S}_{\text{\PM},r,p}(V_{r,p},\lambda_{r,p})
	&\coloneqq
		\lambda_{r,p}
		\left[
		\begin{aligned}
			P_{0,r,p}&f_{\Re,r,p}(V_{r,p})\\
			+jQ_{0,r,p}&f_{\Im,r,p}(V_{r,p})
		\end{aligned}
		\right.
\end{align}
\begin{align}
	f_{\Re,r,p}(V_{r,p})
	&\coloneqq	\alpha_{\Re,r,p}\Abs{\frac{V_{r,p}}{V_{0,r}}}^{2}+\beta_{\Re,r,p}\Abs{\frac{V_{r,p}}{V_{0,r}}}+\gamma_{\Re,r,p}
	\\
	f_{\Im,r,p}(V_{r,p})
	&\coloneqq	\alpha_{\Im,r,p}\Abs{\frac{V_{r,p}}{V_{0,r}}}^{2}+\beta_{\Im,r,p}\Abs{\frac{V_{r,p}}{V_{0,r}}}+\gamma_{\Im,r,p}
\end{align}
where $\lambda$ is a loading factor, $P_{0}, Q_{0}, V_{0}$ are reference values, and $\alpha, \beta, \gamma$ are normalized coefficients.
Furthermore, define
\begin{align}
	\Matrix{S}_{\text{\PM},r}(\Matrix{V}_{r},\Matrix{\lambda}_{r})
	&=	\col_{p\in\Set{P}}\left(\Matrix{S}_{\text{\PM},r}(V_{r,p},\lambda_{r})\right)
	\\
	\Matrix{S}_{\text{\PM}}(\Matrix{V}_{\Set{R}},\Matrix{\lambda})
	&=	\col_{r\in\Set{R}}\left(\Matrix{S}_{\text{\PM},r}(\Matrix{V}_{r},\Matrix{\lambda}_{r})\right)
	\label{eq:pm:pwr}
\end{align}


Zero-injection nodes have zero injected current in all phases.
By consequence, the injected powers are zero:
\begin{equation}
	\mathbf{S}_{\Set{Z}}(\Matrix{V}) = \mathbf{0}
	\label{eq:zero:pwr}
\end{equation}


Finally, note that slack nodes and resource nodes correspond to $V\delta$ buses and $PQ$ buses, respectively.
Due to lack of space, PV buses are not considered in this paper, but their treatment is straightforward \cite{J:PSE:VSA:2005:Zhang}.
\clearpage

\section{Numerical Methods}
\label{sec:meth}


\subsection{Power-Flow Study}
\label{sec:meth:pfs}

The combination of \eqref{eq:grid:pwr}, \eqref{eq:te:pwr}, \eqref{eq:pm:pwr} and \eqref{eq:zero:pwr} yields the \PFE[s] in the form of mismatch equations.
Namely
\begin{equation}
	\Delta\Matrix{S}(\Matrix{V},\Matrix{\lambda})
	\coloneqq
			\Matrix{S}(\Matrix{V})
		-	\left[
			\begin{array}{l}
				\Matrix{S}_{\text{\TE}}(\Matrix{V}_{\Set{S}})\\
				\Matrix{S}_{\text{\PM}}(\Matrix{V}_{\Set{R}},\Matrix{\lambda})\\
				\Matrix{0}_{\Card{\Set{Z}}\cartprod1}\\
			\end{array}
			\right]
	=	\Matrix{0}
	\label{eq:pfs:mismatch}
\end{equation}
Express $\Delta\Matrix{S}(\Matrix{V},\Matrix{\lambda})$ in rectangular and $\Matrix{V}$ in polar coordinates, i.e. $\Delta\Matrix{S}(\Matrix{V},\Matrix{\lambda}) \coloneqq \Delta\Matrix{P}(\Matrix{V},\Matrix{\lambda}) + j\Delta\Matrix{Q}(\Matrix{V},\Matrix{\lambda})$ and $\Matrix{V} \coloneqq \Matrix{E}\angle\Matrix{\theta}$.
For fixed $\Matrix{\lambda}=\Matrix{\lambda}_{\star}$, \eqref{eq:pfs:mismatch} can be reformulated as
\begin{equation}
	\Matrix{f}(\Matrix{E},\Matrix{\theta})
	\coloneqq	\left[
					\begin{array}{l}
						\Delta\Matrix{P}(\Matrix{E}\angle\Matrix{\theta},\Matrix{\lambda}_{\star})\\
						\Delta\Matrix{Q}(\Matrix{E}\angle\Matrix{\theta},\Matrix{\lambda}_{\star})
					\end{array}
					\right]
	=				\Matrix{0}
	\label{eq:pfs:function}
\end{equation}
This equation can be solved with the \NRM in \Alg~\ref{alg:nrm} (see \cite{J:PSE:PFS:1974:Wasley}).
Here, $\Matrix{x}_{\text{\PFS}}\coloneqq[\Matrix{E};\Matrix{\theta}]$ are the unknowns, $\diff_{\Matrix{x}}\Matrix{f}$ is the Jacobian matrix, and $\varepsilon$ the convergence tolerance.

\begin{algorithm}[tb]
	\caption{\NRM for $\Matrix{f}(\Matrix{x})=\Matrix{0}$.}
	\label{alg:nrm}
	\begin{algorithmic}
	\Procedure{\NRM}{$\Matrix{f}(\Matrix{x})$, $\Matrix{x}_{0}$}\Comment{Initial guess $\Matrix{x}_{0}$.}
		\For{$i\geqslant0$}
			\State{$\Delta\Matrix{f} \gets \Matrix{f}(\Matrix{x}_{i})$}
			\If{$\Norm{\Delta\Matrix{f}}\leqslant\varepsilon$}\Comment{Convergence.}
				\State{\textbf{break}}
			\Else\Comment{Correction step.}
				\State{$\Matrix{J} \gets \diff_{\Matrix{x}}\Matrix{f}(\Matrix{x}_{i})$}
				\State{$\Delta\Matrix{x} \gets \text{solve}\left(\Matrix{J}\Delta\Matrix{x}=\Delta\Matrix{f},\Delta\Matrix{x}\right)$}
				\State{$\Matrix{x}_{i+1} \gets \Matrix{x}_{i} - \Delta\Matrix{x}$}
			\EndIf
		\EndFor
		\State{\textbf{return} $\Matrix{x}_{i}$}\Comment{Final solution $\Matrix{x}_{i}$.}
	\EndProcedure
\end{algorithmic}
\end{algorithm}


\subsection{State Estimation}
\label{sec:meth:se}

The slack and resource nodes are equipped with \emph{Phasor Measurement Units} (\PMU[s]), and the zero-injection nodes are treated as virtual measurements \cite{PhD:PSE:SE:2017:Zanni}.
Note that $\Set{S}\cup\Set{R}=\Compl{\Set{Z}}$.
Let $\meas{\Matrix{V}}_{\Compl{\Set{Z}}}$ and $\meas{\Matrix{I}}_{\Compl{\Set{Z}}}$ be the \PMU measurements, $\meas{\Matrix{I}}_{\Set{Z}}\coloneqq\Matrix{0}_{\Card{\Set{Z}}\times1}$ the virtual ones, and $\meas{\Matrix{I}} \coloneqq \col(\meas{\Matrix{I}}_{\Compl{\Set{Z}}},\meas{\Matrix{I}}_{\Set{Z}})$.
Express the states and measurements in rectangular coordinates
\begin{equation}
	\Matrix{x}_{\text{\SE}}
	\coloneqq	\left[
					\begin{array}{l}
						\Real{\Matrix{V}}\\
						\Imag{\Matrix{V}}\\
					\end{array}
					\right]
	,~
	\Matrix{y}
	\coloneqq	\left[
					\begin{array}{l}
						\Re\{\meas{\Matrix{V}}_{\Compl{\Set{Z}}}\}\\
						\Im\{\meas{\Matrix{V}}_{\Compl{\Set{Z}}}\}\\
						\Re\{\meas{\Matrix{I}}\}\\
						\Im\{\meas{\Matrix{I}}\}
					\end{array}
					\right]
\end{equation}
and assume that the measurement noise is white and Gaussian.
This yields a linear measurement model \cite{B:PSE:SE:2016:Paolone}
\begin{equation}
	\mathbf{y} = \mathbf{C}\mathbf{x}_{\text{\SE}} + \mathbf{v},~
	\mathbf{v} \sim \mathbf{N}(\mathbf{0},\mathbf{R})
	\label{eq:se:function}
\end{equation}
where $\mathbf{N}(\mathbf{0},\mathbf{R})$ is the multivariate normal distribution with mean vector $\Matrix{0}$ and covariance matrix $\Matrix{R}$.
$\Matrix{C}$ is built as follows
\begin{equation}
	\Matrix{C}
	=	\left[
		\begin{array}{ll}
			\Matrix{\Gamma}	&\phantom{-}\Matrix{0}\\
			\Matrix{0}				&\phantom{-}\Matrix{\Gamma}\\
			\Matrix{G}				&-\Matrix{B}\\
			\Matrix{B}				&\phantom{-}\Matrix{G}\\
		\end{array}
		\right]
\end{equation}
where $\Matrix{\Gamma}$ is the indicator function for the voltages $\Matrix{V}_{\Compl{Z}}$
\begin{equation}
	\Matrix{\Gamma}\in\mathbb{R}^{\Card{\Compl{\Set{Z}}}\times\Card{\Set{N}}},~
	\Gamma_{mn}
	\coloneqq	\left\{
					\begin{array}{ll}
						1	&\text{if}~m=n\\
						0	&\text{otherwise}
					\end{array}
					\right.
\end{equation}
and $\Matrix{G}\coloneqq\Real{\Matrix{Y}}$, $\Matrix{B}\coloneqq\Imag{\Matrix{Y}}$.
The linear \WLSR in \Alg~\ref{alg:lsr} yields an estimate $\widehat{\Matrix{x}}_{\text{\SE}}$ with minimum squared error (see \cite{B:PSE:SE:2016:Paolone}).

\begin{algorithm}[tb]
	\caption{Linear \WLSR for $\Matrix{y}=\Matrix{C}\Matrix{x} + \Matrix{v}$, $\Matrix{v}\sim\Matrix{N}(\Matrix{0},\Matrix{R})$.}
	\label{alg:lsr}
	\begin{algorithmic}
	\Procedure{L\WLSR}{$\Matrix{y}, \Matrix{C}, \Matrix{R}$}
		\State{$\Matrix{G} \gets \Tran{\Matrix{C}}\Inv{\Matrix{R}}\Matrix{C}$}\Comment{Gain matrix.}
		\State{$\estim{\Matrix{x}} \gets \Inv{\Matrix{G}}\Tran{\Matrix{C}}\Inv{\Matrix{R}}\Matrix{y}$}
		\State{\textbf{return} $\estim{\Matrix{x}}$}
	\EndProcedure
\end{algorithmic}
\end{algorithm}


\subsection{Voltage Stability Assessment}
\label{sec:meth:vsa}

Suppose that $\Matrix{\lambda}$ follows the trajectory $\Matrix{\lambda}(\lambda) \coloneqq \Matrix{\lambda}_{0} + \lambda\cdot\Matrix{t}$ with origin $\Matrix{\lambda}_{0}$ and direction $\Matrix{t}$.
Define $\Matrix{g}(\Matrix{E},\Matrix{\theta},\lambda)$ as the analogon of $\Matrix{f}(\Matrix{E},\Matrix{\theta})$, which includes the trajectory $\Matrix{\lambda}(\lambda)$.
Namely
\begin{equation}
	\Matrix{g}(\Matrix{E},\Matrix{\theta},\lambda)
	\coloneqq	\left[
					\begin{array}{l}
						\Delta\Matrix{P}(\Matrix{E}\angle\Matrix{\theta},\Matrix{\lambda}(\lambda))\\
						\Delta\Matrix{Q}(\Matrix{E}\angle\Matrix{\theta},\Matrix{\lambda}(\lambda))
					\end{array}
					\right]
	=				\Matrix{0}
	\label{eq:vsa:function}
\end{equation}
The objective of \VSA is to find the maximum $\lambda$ for which the above-stated equations remains solvable.
That is
\begin{equation}
	\max\lambda~\text{s.t.}~\Matrix{g}(\Matrix{E},\Matrix{\theta},\lambda) = \Matrix{0}
\end{equation}
\CPF methods solve this optimization problem by producing a continuum of solutions of \eqref{eq:vsa:function}.
For this, the homotopy \CM in \Alg~\ref{alg:hcm} is used (see \cite{J:PSE:VSA:1995:Chiang,J:PSE:VSA:2014:Sheng}).
Note that $\Matrix{x}_{\text{\VSA}}\coloneqq[\Matrix{E};\Matrix{\theta}]$,~$\xi\coloneqq\lambda$ are the unknowns, and $\sigma$ is the step size used for continuation.
The \CM employs a predictor to calculate guesses $\Matrix{x}^{-}_{k+1}$, $\xi^{-}_{k+1}$ of the next solutions in the continuum, and a corrector to find the actual values $\Matrix{x}^{+}_{k+1}$, $\xi^{+}_{k+1}$.
The predictor is based on the tangent method, and the corrector on the \NRM.

\begin{algorithm}[tb]
	\caption{Homotopy \CM for $\max\xi$ s.t. $\Matrix{g}(\Matrix{x},\xi)=\Matrix{0}$.}
	\label{alg:hcm}
	\begin{algorithmic}
	\Procedure{H\CM}{$\Matrix{g}(\Matrix{x},\xi)$, $\Matrix{x}_{0}$, $\xi_{0}$}\Comment{Starting point $\Matrix{x}_{0},\xi_{0}$.}
		\For{k$\geqslant0$}
		\State{\# Predictor (tangent method)}
		\State{$d\Matrix{x} \gets \text{solve}\left(\diff_{\Matrix{x}}\Matrix{g}(\Matrix{x}_{k},\xi_{k})d\Matrix{x}=-\diff_{\xi}\Matrix{g}(\Matrix{x}_{k},\xi_{k}),d\Matrix{x}\right)$}
		\State{%
			$
			\left[
			\begin{array}{l}
				\Matrix{x}^{-}_{k+1}\\
				\xi^{-}_{k+1}
			\end{array}
			\right]
			=		\left[
					\begin{array}{l}
						\Matrix{x}_{k}\\
						\xi_{k}
					\end{array}
					\right]
				+	\sigma\left(\frac{1}{\sqrt{\Norm{d\Matrix{x}}^{2}+1}}
					\left[
					\begin{array}{c}
						d\Matrix{x}\\
						1
					\end{array}
					\right]\right)
			$%
		}
		\vspace{3pt}
		\State{\# Corrector (\NRM)}
		\vspace{1pt}
		\State{%
			$
			\Matrix{h}([\Matrix{x};\xi])
			\coloneqq	\left[\hspace{-3pt}
							\begin{array}{l}
								\Matrix{g}(\Matrix{x},\xi)\\
								\Norm{\Matrix{x}-\Matrix{x}_{k}}^{2} + (\xi-\xi_{k})^{2} - \sigma^{2}
							\end{array}
							\hspace{-3pt}\right]
			$%
		}
		\State{$[\Matrix{x}^{+}_{k+1};\xi^{+}_{k+1}] \gets \text{NRM}\left(\Matrix{h}([\Matrix{x};\xi]), [\Matrix{x}^{-}_{k+1};\xi^{-}_{k+1}]\right)$}
		\State{%
			$
			\left[
			\begin{array}{l}
				\Matrix{x}_{k+1}\\
				\xi_{k+1}
			\end{array}
			\right]
			\gets	\left[
					\begin{array}{l}
						\Matrix{x}^{+}_{k+1}\\
						\xi^{+}_{k+1}
					\end{array}
					\right]
			$%
		}
		\vspace{2pt}
		\If{$\sign(\xi_{k+1}-\xi_{k})\leqslant0$}\Comment{Maximum $\xi$.}
			\State{\textbf{break}}
		\EndIf
		\EndFor
		\State{\textbf{return} $\{\Matrix{x}_{k}, \xi_{k}\}$}\Comment{Continuum of solutions $\{\Matrix{x}_{k},\xi_{k}\}$.}
	\EndProcedure
\end{algorithmic}
\end{algorithm}
\clearpage

\section{Performance Assessment}
\label{sec:perf}


\subsection{Test System \& Simulation Setup}

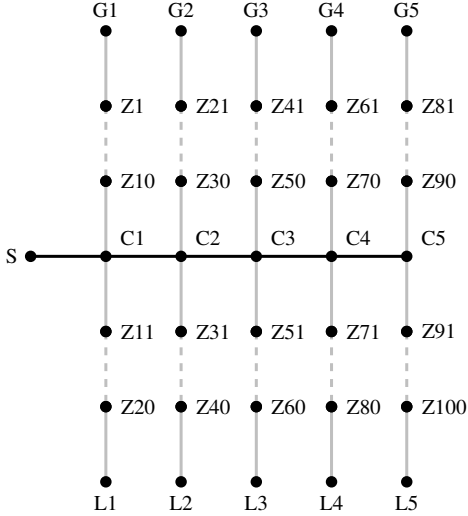
\begin{figure}[t]
	\centering
	\tikzstyle{major}=[black,line width=0.4mm]
\tikzstyle{minor}=[lightgray,line width=0.4mm]

\begin{circuitikz}
	\footnotesize
	
	\def\dx{1.0}
	\def\dy{1.0}
	\def\LabelSeparation{0.25}
	
	
	\coordinate (S) at (0,1);	
	
	\coordinate (C1) at ($(S)+(\dx,0)$);
	\coordinate (C2) at ($(C1)+(\dx,0)$);
	\coordinate (C3) at ($(C2)+(\dx,0)$);
	\coordinate (C4) at ($(C3)+(\dx,0)$);
	\coordinate (C5) at ($(C4)+(\dx,0)$);
	
	\coordinate (G1) at ($(C1)+3*(0,\dy)$);
	\coordinate (G2) at ($(C2)+3*(0,\dy)$);
	\coordinate (G3) at ($(C3)+3*(0,\dy)$);
	\coordinate (G4) at ($(C4)+3*(0,\dy)$);
	\coordinate (G5) at ($(C5)+3*(0,\dy)$);
	
	\coordinate (L1) at ($(C1)-3*(0,\dy)$);
	\coordinate (L2) at ($(C2)-3*(0,\dy)$);
	\coordinate (L3) at ($(C3)-3*(0,\dy)$);
	\coordinate (L4) at ($(C4)-3*(0,\dy)$);
	\coordinate (L5) at ($(C5)-3*(0,\dy)$);
	
	\coordinate (Z1) at ($(G1)-(0,\dy)$);
	\coordinate (Z10) at ($(C1)+(0,\dy)$);
	\coordinate (Z11) at ($(C1)-(0,\dy)$);
	\coordinate (Z20) at ($(L1)+(0,\dy)$);
	\coordinate (Z21) at ($(G2)-(0,\dy)$);
	\coordinate (Z30) at ($(C2)+(0,\dy)$);
	\coordinate (Z31) at ($(C2)-(0,\dy)$);
	\coordinate (Z40) at ($(L2)+(0,\dy)$);
	\coordinate (Z41) at ($(G3)-(0,\dy)$);
	\coordinate (Z50) at ($(C3)+(0,\dy)$);
	\coordinate (Z51) at ($(C3)-(0,\dy)$);
	\coordinate (Z60) at ($(L3)+(0,\dy)$);
	\coordinate (Z61) at ($(G4)-(0,\dy)$);
	\coordinate (Z70) at ($(C4)+(0,\dy)$);
	\coordinate (Z71) at ($(C4)-(0,\dy)$);
	\coordinate (Z80) at ($(L4)+(0,\dy)$);
	\coordinate (Z81) at ($(G5)-(0,\dy)$);
	\coordinate (Z90) at ($(C5)+(0,\dy)$);
	\coordinate (Z91) at ($(C5)-(0,\dy)$);
	\coordinate (Z100) at ($(L5)+(0,\dy)$);
	
	
	\draw[minor,dashed] (Z1) to (Z10);
	\draw[minor,dashed] (Z11) to (Z20);
	\draw[minor] (G1) [short,*-*] to (Z1) to[open] (Z10) to[short,*-] (C1) to[short,-*] (Z11) to[open] (Z20) to[short,*-*] (L1);
	
	\draw[minor,dashed] (Z21) to (Z30);
	\draw[minor,dashed] (Z31) to (Z40);
	\draw[minor] (G2) [short,*-*] to (Z21) to[open] (Z30) to[short,*-] (C2) to[short,-*] (Z31) to[open] (Z40) to[short,*-*] (L2);
	
	\draw[minor,dashed] (Z41) to (Z50);
	\draw[minor,dashed] (Z51) to (Z60);
	\draw[minor] (G3) [short,*-*] to (Z41) to[open] (Z50) to[short,*-] (C3) to[short,-*] (Z51) to[open] (Z60) to[short,*-*] (L3);
	
	\draw[minor,dashed] (Z61) to (Z70);
	\draw[minor,dashed] (Z71) to (Z80);
	\draw[minor] (G4) [short,*-*] to (Z61) to[open] (Z70) to[short,*-] (C4) to[short,-*] (Z71) to[open] (Z80) to[short,*-*] (L4);
	
	\draw[minor,dashed] (Z81) to (Z90);
	\draw[minor,dashed] (Z91) to (Z100);
	\draw[minor] (G5) [short,*-*] to (Z81) to[open] (Z90) to[short,*-] (C5) to[short,-*] (Z91) to[open] (Z100) to[short,*-*] (L5);
	
	\draw[major] (S) to[short,*-] (C1) to[short,*-] (C2) to[short,*-] (C3) to[short,*-] (C4) to[short,*-*] (C5);
	
	
	\node at ($(S)-\LabelSeparation*(1,0)$) {S};
	
	\node at ($(C1)+\LabelSeparation*(2,1)$) {C1\hphantom{00}};
	\node at ($(C2)+\LabelSeparation*(2,1)$) {C2\hphantom{00}};
	\node at ($(C3)+\LabelSeparation*(2,1)$) {C3\hphantom{00}};
	\node at ($(C4)+\LabelSeparation*(2,1)$) {C4\hphantom{00}};
	\node at ($(C5)+\LabelSeparation*(2,1)$) {C5\hphantom{00}};
	
	\node at ($(G1)+\LabelSeparation*(0,1.1)$) {G1};
	\node at ($(G2)+\LabelSeparation*(0,1.1)$) {G2};
	\node at ($(G3)+\LabelSeparation*(0,1.1)$) {G3};
	\node at ($(G4)+\LabelSeparation*(0,1.1)$) {G4};
	\node at ($(G5)+\LabelSeparation*(0,1.1)$) {G5};
	
	\node at ($(L1)-\LabelSeparation*(0,1.1)$) {L1};
	\node at ($(L2)-\LabelSeparation*(0,1.1)$) {L2};
	\node at ($(L3)-\LabelSeparation*(0,1.1)$) {L3};
	\node at ($(L4)-\LabelSeparation*(0,1.1)$) {L4};
	\node at ($(L5)-\LabelSeparation*(0,1.1)$) {L5};
	
	\node at ($(Z1)+\LabelSeparation*(2,0)$) {Z1\phantom{00}};
	\node at ($(Z10)+\LabelSeparation*(2,0)$) {Z10\hphantom{0}};
	\node at ($(Z11)+\LabelSeparation*(2,0)$) {Z11\hphantom{0}};
	\node at ($(Z20)+\LabelSeparation*(2,0)$) {Z20\hphantom{0}};
	\node at ($(Z21)+\LabelSeparation*(2,0)$) {Z21\hphantom{0}};
	\node at ($(Z30)+\LabelSeparation*(2,0)$) {Z30\hphantom{0}};
	\node at ($(Z31)+\LabelSeparation*(2,0)$) {Z31\hphantom{0}};
	\node at ($(Z40)+\LabelSeparation*(2,0)$) {Z40\hphantom{0}};
	\node at ($(Z41)+\LabelSeparation*(2,0)$) {Z41\hphantom{0}};
	\node at ($(Z50)+\LabelSeparation*(2,0)$) {Z50\hphantom{0}};
	\node at ($(Z51)+\LabelSeparation*(2,0)$) {Z51\hphantom{0}};
	\node at ($(Z60)+\LabelSeparation*(2,0)$) {Z60\hphantom{0}};
	\node at ($(Z61)+\LabelSeparation*(2,0)$) {Z61\hphantom{0}};
	\node at ($(Z70)+\LabelSeparation*(2,0)$) {Z70\hphantom{0}};
	\node at ($(Z71)+\LabelSeparation*(2,0)$) {Z71\hphantom{0}};
	\node at ($(Z80)+\LabelSeparation*(2,0)$) {Z80\hphantom{0}};
	\node at ($(Z81)+\LabelSeparation*(2,0)$) {Z81\hphantom{0}};
	\node at ($(Z90)+\LabelSeparation*(2,0)$) {Z90\hphantom{0}};
	\node at ($(Z91)+\LabelSeparation*(2,0)$) {Z91\phantom{0}};
	\node at ($(Z100)+\LabelSeparation*(2,0)$) {Z100};
	
\end{circuitikz}
	\caption
	{%
		Single-line diagram of the test system.
		The nodes comprise one slack node $\Set{S}=\{\text{S}\}$, 15 resource nodes $\Set{R}=\{\text{G1--G5}\}\cup\{\text{L1--L5}\}\cup\{\text{C1--C5}\}$, and 100 zero-injection nodes $\Set{Z}=\{\text{Z1--Z100}\}$. 
		The lines are untransposed, and of type \IEEE-300 (black) and \IEEE-301 (gray), respectively (see \cite{R:PSE:PFS:2004:IEEE}).
	}
	\label{fig:schematic}
\end{figure}

\begin{figure}[!ht]
	\centering
	
	\includegraphics[width=\linewidth]{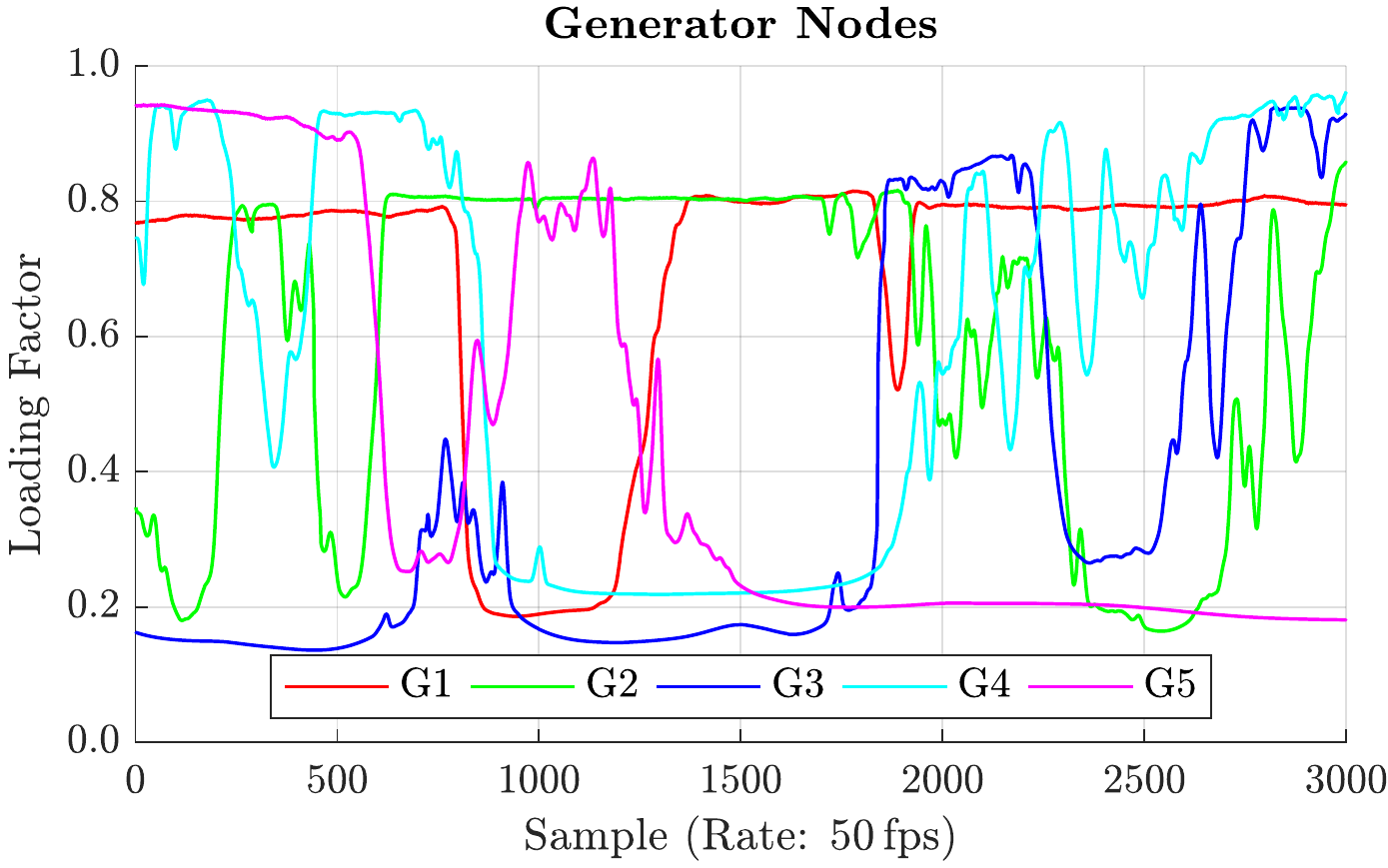}
	
	\vspace{1mm}
	
	\includegraphics[width=\linewidth]{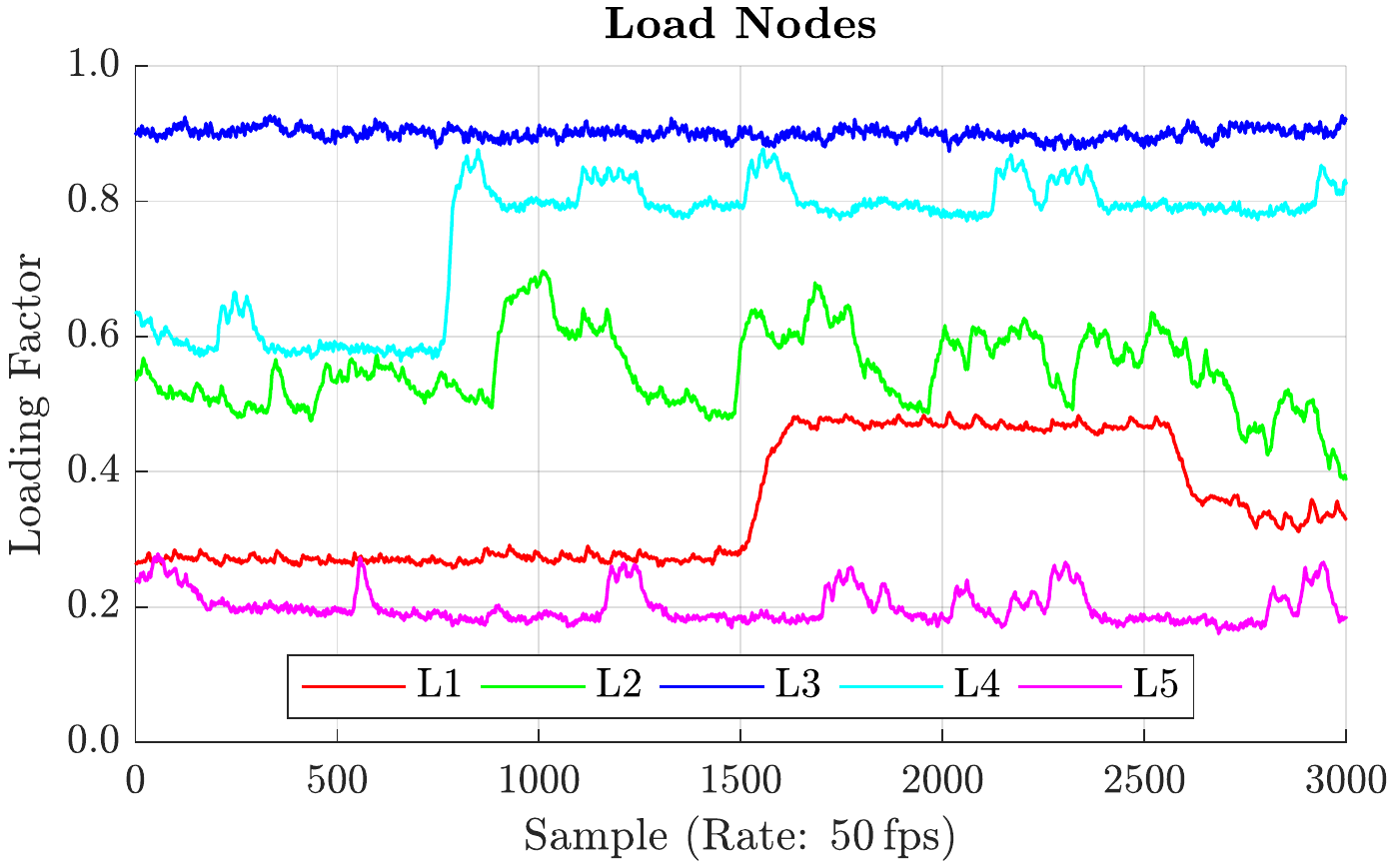}
	
	\caption{Profiles of the loading factors at the generator and load nodes.}
	\label{fig:profiles}
\end{figure}

The numerical methods discussed in \Sec~\ref{sec:meth} are applied to an unbalanced three-phase power grid. %
The single-line diagram of the test system is depicted in \Fig~\ref{fig:schematic}.
%
%
The grid is composed of untransposed overhead lines of 5\,km length each, which are configured according to the codes \IEEE-300 and \IEEE-301 from \cite{R:PSE:PFS:2004:IEEE}.
%
%
The system has 1 slack node (S), 15 resource nodes (i.e., generator nodes G1--G5, load nodes L1--L5, and compensator nodes C1--C5), and 100 zero-injection nodes (Z1--Z100).
The slack node has a short-circuit power of $S_{\text{sc}}=100$\,\text{MVA}.
Its \TE consists of a positive-sequence voltage source rated at nominal voltage, and a diagonal compound impedance matrix with equal diagonal entries, for which $R/X=0.1$.
The \PM[s] of the resource nodes are described by the parameters listed in \Tab[s]~\ref{tab:params:refvals}--\ref{tab:params:coeffs}.
Notably, the load coefficients are derived from real-world data \cite{J:PSE:PFS:1988:Price}.
The loading factors are considered to be equal in all phases of a given node (i.e., $\lambda_{r,p}=\lambda_{r}$ $\forall r\in\Set{R}$).
For the generator and load nodes, the profiles shown in \Fig~\ref{fig:profiles} are used.
These profiles are derived from power measurements recorded in the medium-voltage grid of the \EPFL campus \cite{J:PSE:SE:2015:Pignati}.
For the compensator nodes, the loading factors are equal to 1.


\KR is performed in 11 steps, which are numbered as 0--10.
Step~0 denotes the base case, in which all nodes are considered, step~1 corresponds to the reduction of Z91-Z100, step~2 to the reduction of Z81-Z100, and so forth.
The electrical quantities are expressed in \emph{per unit} (pu) of the per-unit system specified by $P_{b}=10$\,MW and $V_{b}=24.9$\,kV~phase-to-phase.


The numerical methods are coded in MATLAB (R2018a), and run on a MacBook Pro (mid 2014, 2.5\,GHz Intel Core i7, 16\,GB 1600\,MHz DDR3 RAM).
The code is based exclusively on dense linear algebra routines, since \KR reduces the sparsity of the compound admittance matrix.

\begin{table}[!ht]
	\centering
	\caption{Reference voltages and powers of the resource nodes.}
	\label{tab:params:refvals}
	{
\renewcommand{\arraystretch}{1.2}
\begin{tabular}{ccccc}
	\hline
		Node	
	&	$V_{0}$	&$P_{0,A}$,~$P_{0,B}$,~$P_{0,C}$
	&	$Q_{0,A}$,~$Q_{0,B}$,~$Q_{0,C}$
	&	Type
	\\
	&	(kV)
	&	(kW)
	&	(kVAR)\\			
	\hline
		G1
	&	14.4
	&	\phantom{$-$}100,~\phantom{$-$}100,~\phantom{$-$}100
	&	\phantom{00$-$}0,~\phantom{00$-$}0,~\phantom{00$-$}0
	&	G
	\\
		G2
	&	14.4
	&	\phantom{$-$}240,~\phantom{$-$}160,~\phantom{0$-$}80
	&	\phantom{00$-$}0,~\phantom{00$-$}0,~\phantom{00$-$}0
	&	G
	\\
		G3
	&	14.4
	&	\phantom{$-$}110,~\phantom{$-$}190,~\phantom{$-$}150
	&	\phantom{00$-$}0,~\phantom{00$-$}0,~\phantom{00$-$}0
	&	G
	\\
		G4
	&	14.4
	&	\phantom{0$-$}60,~\phantom{$-$}120,~\phantom{$-$}180
	&	\phantom{00$-$}0,~\phantom{00$-$}0,~\phantom{00$-$}0
	&	G
	\\
		G5
	&	14.4
	&	\phantom{$-$}150,~\phantom{$-$}150,~\phantom{$-$}150
	&	\phantom{00$-$}0,~\phantom{00$-$}0,~\phantom{00$-$}0
	&	G
	\\
	\hline
		L1
	&	14.4
	&	$-$200,~$-$200,~$-$200
	&	\phantom{0}$-$40,~\phantom{0}$-$40,~\phantom{0}$-$40
	&	L
	\\
		L2	
	&	14.4
	&	\phantom{0}$-$90,~$-$110,~$-$130
	&	\phantom{00}$-$9,~\phantom{0}$-$11,~\phantom{0}$-$13
	&	L
	\\
		L3	
	&	14.4
	&	$-$120,~$-$150,~$-$180
	&	\phantom{0}$-$12,~\phantom{0}$-$15,~\phantom{0}$-$18
	&	L
	\\
		L4
	&	14.4
	&	$-$100,~$-$120,~$-$140
	&	\phantom{0}$-$10,~\phantom{0}$-$12,~\phantom{0}$-$14
	&	L
	\\
		L5
	&	14.4
	&	$-$250,~$-$250,~$-$250
	&	\phantom{0}$-$50,~\phantom{0}$-$50,~\phantom{0}$-$50
	&	L
	\\
	\hline
		C1
	&	14.4
	&	\phantom{00$-$}0,~\phantom{00$-$}0,~\phantom{00$-$}0
	&	\phantom{0$-$}20,~\phantom{0$-$}20,~\phantom{0$-$}20
	&	C
	\\
		C2
	&	14.4
	&	\phantom{00$-$}0,~\phantom{00$-$}0,~\phantom{00$-$}0
	&	\phantom{0$-$}30,~\phantom{0$-$}30,~\phantom{0$-$}30
	&	C
	\\
		C3
	&	14.4
	&	\phantom{00$-$}0,~\phantom{00$-$}0,~\phantom{00$-$}0
	&	\phantom{0$-$}40,~\phantom{0$-$}40,~\phantom{0$-$}40
	&	C
	\\
		C4
	&	14.4
	&	\phantom{00$-$}0,~\phantom{00$-$}0,~\phantom{00$-$}0
	&	\phantom{0$-$}50,~\phantom{0$-$}50,~\phantom{0$-$}50
	&	C
	\\
		C5
	&	14.4
	&	\phantom{00$-$}0,~\phantom{00$-$}0,~\phantom{00$-$}0
	&	\phantom{0$-$}60,~\phantom{0$-$}60,~\phantom{0$-$}60
	&	C
	\\
	\hline
\end{tabular}
}	
\end{table}

\begin{table}[!ht]
	\centering
	\caption{Polynomial coefficients of the resource nodes.}
	\label{tab:params:coeffs}
	{
\renewcommand{\arraystretch}{1.2}
\begin{tabular}{ccc}
	\hline
		Type
	&	$\alpha_{\Re}$,~$\beta_{\Re}$,~$\gamma_{\Re}$
	&	$\alpha_{\Im}$,~$\beta_{\Im}$,~$\gamma_{\Im}$
	\\
	\hline
		G
	&	\hphantom{$-$}0.000,~\hphantom{$-$}0.000,~\hphantom{$-$}1.000
	&	\hphantom{$-$}0.000,~\hphantom{$-$}0.000,~\hphantom{$-$}1.000
	\\
		L				
	&	$-$0.067,~\hphantom{$-$}0.251,~\hphantom{$-$}0.816
	&	\hphantom{$-$}1.064,~$-$0.088,~\hphantom{$-$}0.025
	\\
		C
	&	\hphantom{$-$}1.000,~\hphantom{$-$}0.000,~\hphantom{$-$}0.000
	&	\hphantom{$-$}1.000,~\hphantom{$-$}0.000,~\hphantom{$-$}0.000
	\\
	\hline
\end{tabular}
}	
\end{table}

\clearpage


\begin{table}[t]
	\centering
	\caption{Condition number of the power-flow Jacobian matrix $\Matrix{J}$.}
	\label{tab:pfs:jacobian}
	{
\renewcommand{\arraystretch}{1.2}

\begin{tabular}{ccccccc}
	\hline
		Reduction Step
	&	0
	&	1
	&	2
	&	3
	&	4
	&	5
	\\
	\hline
		$\operatorname{cond}(\Matrix{J})$
	&	6.9E3
	&	6.0E3
	&	5.2E3
	&	4.5E3
	&	3.7E3
	&	3.2E3
	\\
	\hline
\end{tabular}


\vspace{1mm}

\begin{tabular}{ccccccc}
	\hline
		Reduction Step
	&	6
	&	7
	&	8
	&	9
	&	10
	&	\phantom{11}
	\\
	\hline
		$\operatorname{cond}(\Matrix{J})$
	&	2.5E3
	&	2.1E3
	&	1.6E3
	&	8.5E2
	&	4.9E2
	&	\phantom{0.0E0}
	\\
	\hline
\end{tabular}


}	
\end{table}

\begin{table}[t]
	\centering
	\caption{Condition number of the estimator gain matrix $\Matrix{G}$.}
	\label{tab:se:gain}
	{
\renewcommand{\arraystretch}{1.2}

\begin{tabular}{ccccccc}
	\hline
		Reduction Step
	&	0
	&	1
	&	2
	&	3
	&	4
	&	5
	\\
	\hline
		$\operatorname{cond}(\Matrix{G})$
	&	8.0E9
	&	7.3E9
	&	6.8E9
	&	6.2E9
	&	5.7E9
	&	5.2E9
	\\
	\hline
\end{tabular}

\vspace{1mm}

\begin{tabular}{ccccccc}
	\hline
		Reduction Step
	&	6
	&	7
	&	8
	&	9
	&	10
	&	\phantom{11}
	\\
	\hline
		$\operatorname{cond}(\Matrix{G})$
	&	4.7E9
	&	4.1E9
	&	3.4E9
	&	4.4E8
	&	3.3E4
	&	\phantom{0.0E0}
	\\
	\hline
\end{tabular}

}


\end{table}


\subsection{Power-Flow Study}

\begin{figure}[!ht]
	\centering	
	\includegraphics[width=\linewidth]{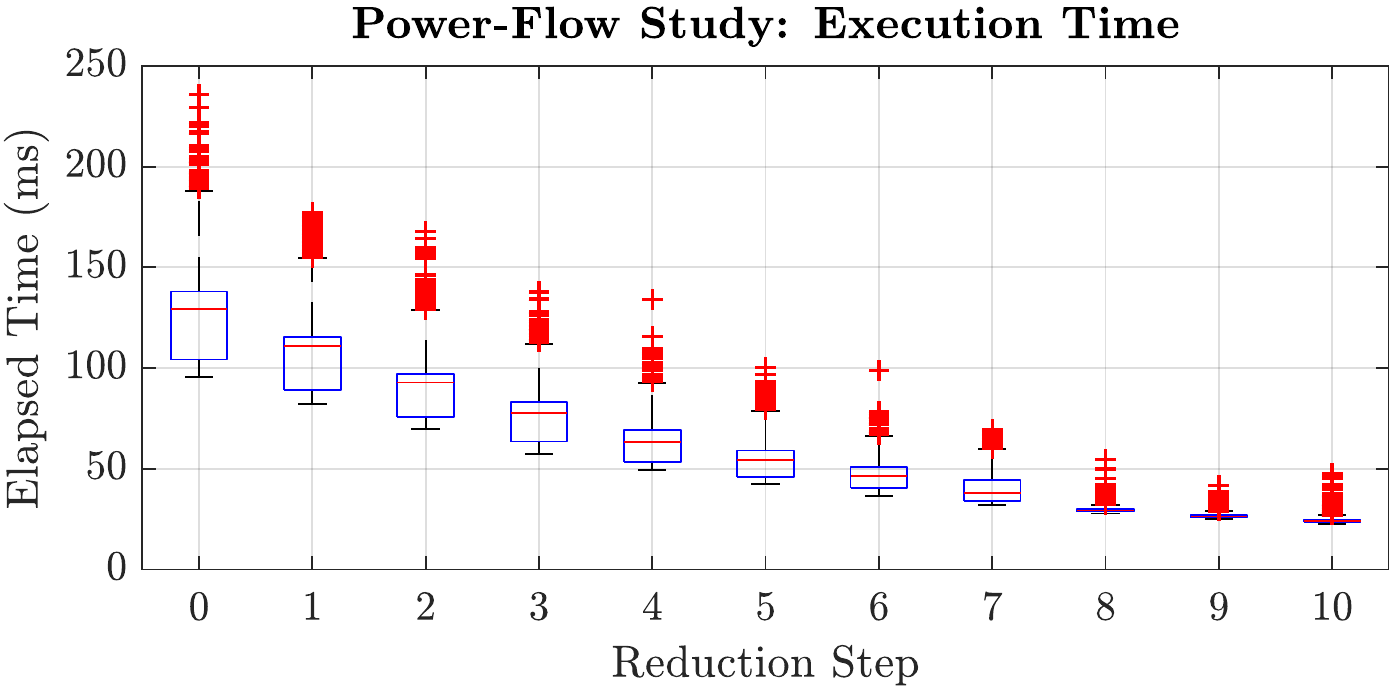}
	\caption{Execution time of the \NRM used for \PFS.}
	\label{fig:pfs:time}
\end{figure}

For the \NRM, the convergence tolerance is set to $\varepsilon=10^{-8}$, and positive-sequence voltage phasors of magnitude 1 are used as initial points.
Convergence is reached after 4--5 iterations.
The key performance indicators of the \NRM are the condition number of the Jacobian matrix (\Tab~\ref{tab:pfs:jacobian}) and the execution time (\Fig~\ref{fig:pfs:time}).
Through steps 0--10 of \KR (i.e., from the original to the fully reduced system), the condition number improves by a factor~of~14, and the median execution time by a factor~of~5.


\subsection{State Estimation}

For \SE, it is supposed that all slack and resource nodes are equipped with \emph{Phasor Measurement Units} (\PMU[s]), which measure the phase-to-ground voltages and injected currents in all phases.
Moreover, all remaining zero-injection nodes are treated as virtual measurements.
The \PMU[s] have a \emph{Full-Scale Range} (\FSR) of 20\,kV (\RMS) for the voltage phasors and 100\,A (\RMS) for the current phasors.
The standard deviations of the measurement noise are $10^{-3}$\,pu (w.r.t. the \FSR) for the magnitudes and $1.5\cdot10^{-3}$\,rad for the angles.
These values are typical for class 0.1 of voltage/current instrument transformers \cite{Std:IEC:61869-2,Std:IEC:61869-3}.
For the virtual measurements, these standard deviations are set 100 times smaller.
The \PMU[s] are emulated by polluting the voltage phasors obtained in the \PFS with suitably scaled white Gaussian noise.

\begin{figure}[!ht]
	\centering	
	\includegraphics[width=\linewidth]{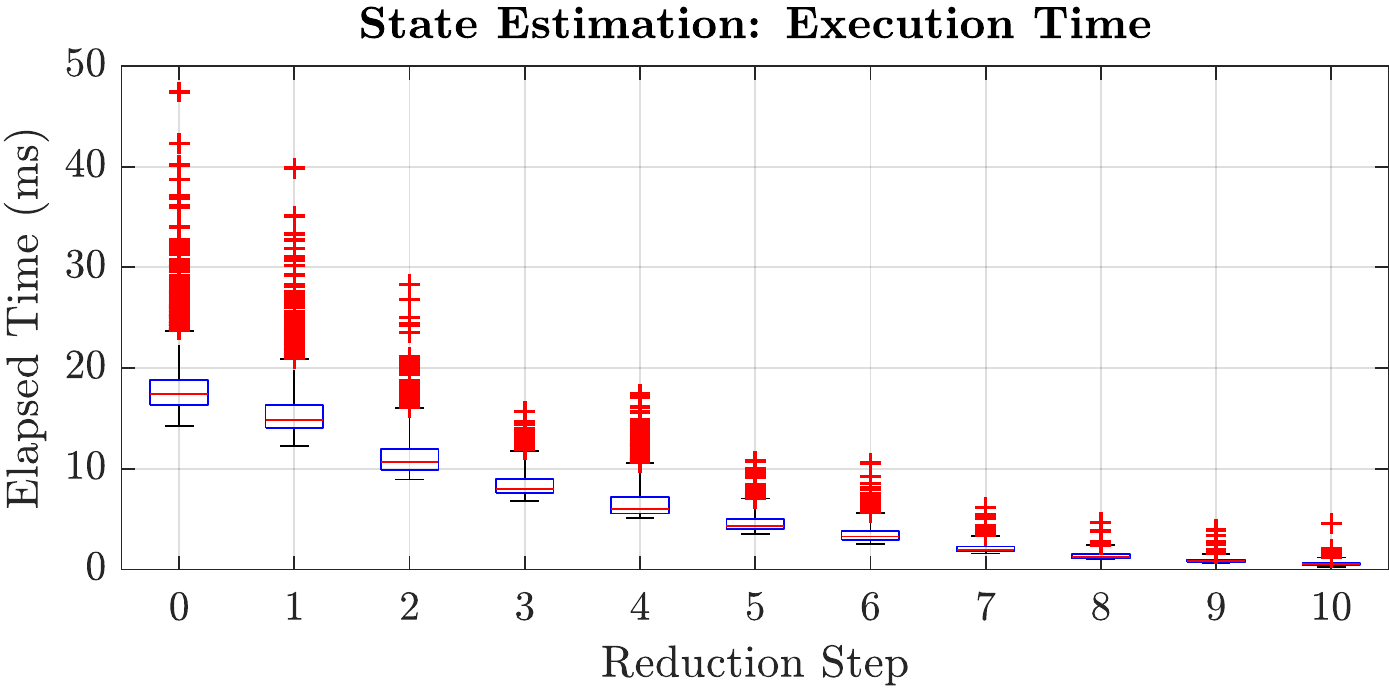}
	\caption{Execution time of the linear \WLSR used for \SE.}
	\label{fig:se:time}
\end{figure}

The performance indicators of the \WLSR are the condition number of the gain matrix (\Tab~\ref{tab:se:gain}) and the execution time (\Fig~\ref{fig:se:time}).
From step 0 to step 10 of \KR, the condition number improves by 5 orders of magnitude (this number depends on the assumed standard deviations of the virtual measurements, see \cite{PhD:PSE:SE:2017:Zanni}), and the median execution time by a factor of 40.


\subsection{Voltage Stability Assessment}

\begin{figure}[!ht]
	\centering	
	\includegraphics[width=\linewidth]{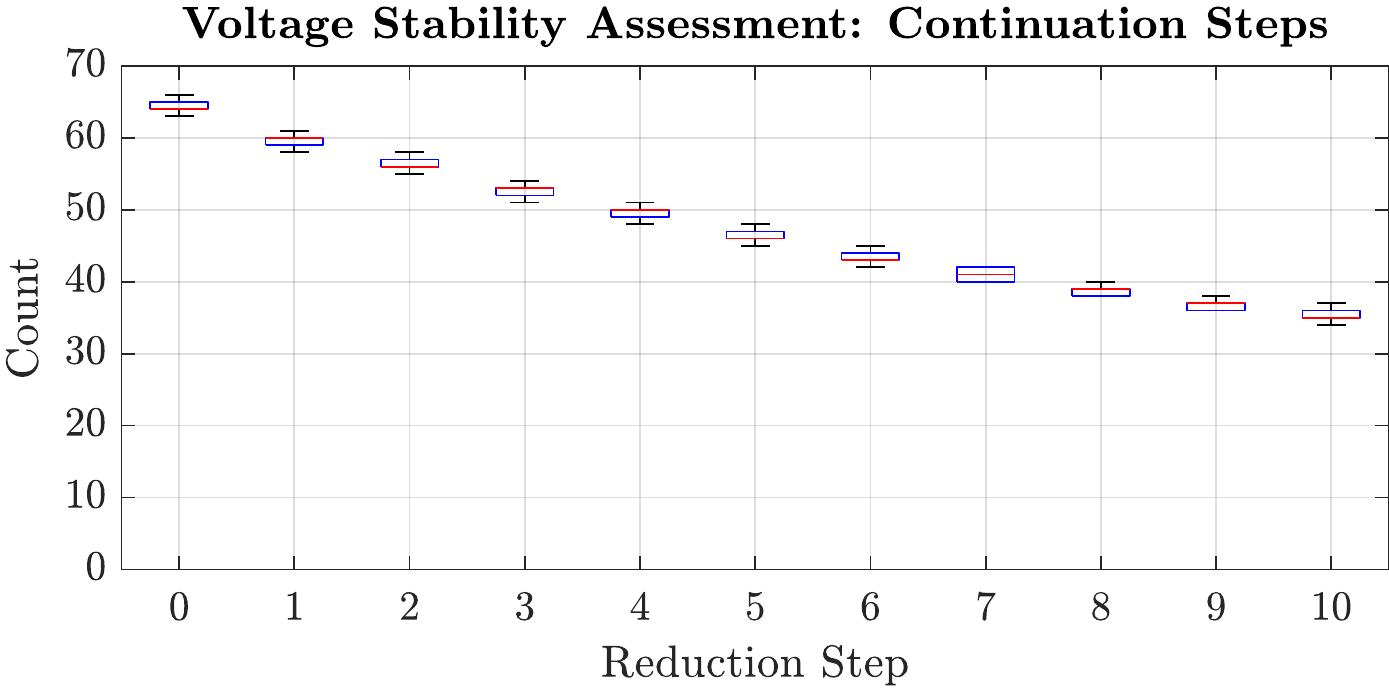}
	\caption{Number of steps of the homotopy \CM used for \VSA.}
	\label{fig:vsa:steps}
\end{figure}

\begin{figure}[!ht]
	\centering	
	\includegraphics[width=\linewidth]{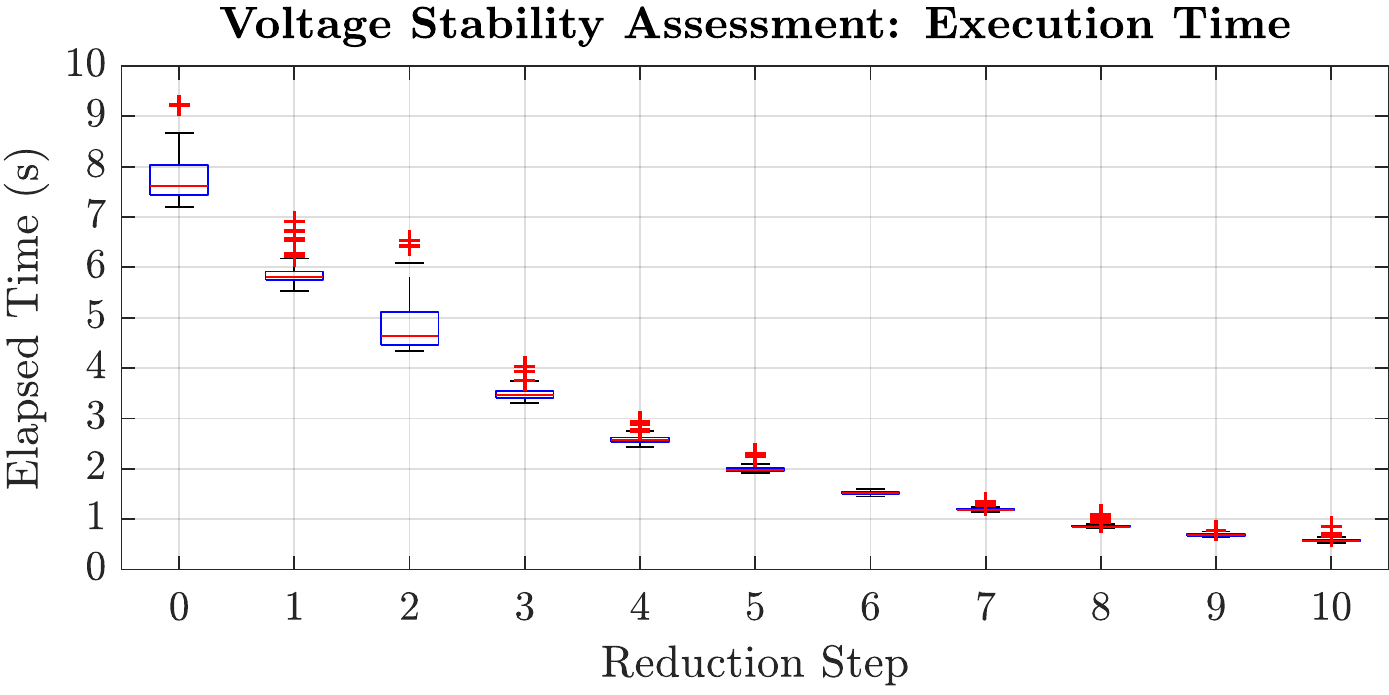}
	\caption{Execution time of the homotopy \CM used for \VSA.}
	\label{fig:vsa:time}
\end{figure}

For the \VSA, only the loading factors of the load nodes are varied.
The \CM uses a convergence tolerance of $\varepsilon=10^{-8}$ and a step size of $\sigma=10^{-1}$.
The key performance indicators of the H\CM are the number of continuation steps (\Fig~\ref{fig:vsa:steps}) and the execution time (\Fig~\ref{fig:vsa:time}).
Through the application of \KR, the number of continuation steps is approximately halved, and the median execution time is reduced by a factor of 10.


\clearpage
\clearpage

\section{Conclusion}
\label{sec:concl}

This paper examined the impact of \KR on the performance of state-of-the-art numerical methods applied to the analysis of unbalanced polyphase power systems.
Namely, the classical applications \PFS, \SE, and \VSA were considered.
To this end, the \NRM, linear \WLSR, and homotopy \CM were implemented in MATLAB.
The impact of \KR on the performance of these methods was assessed using a reproducible test system, from which the zero-injection nodes were successively eliminated.
Through the application of \KR, the condition number of the power-flow Jacobian matrix was improved by a factor of 14, and the condition number of the estimator gain matrix by 5 orders of magnitude.
The median execution times of the \NRM, \WLSR, and \CM were reduced by factors of 5, 40, and 10.
These results confirm the applicability and usefulness of \KR for the analysis of unbalanced polyphase power systems.



\section*{Acknowledgements}

This work was supported by the Swiss National Science Foundation via the National Research Programme NRP-70 ``Energy Turnaround'' (project name ``Commelec'').


\bibliographystyle{IEEEtran}
\bibliography{Bibliography}

\begin{thebibliography}{10}
\providecommand{\url}[1]{#1}
\csname url@samestyle\endcsname
\providecommand{\newblock}{\relax}
\providecommand{\bibinfo}[2]{#2}
\providecommand{\BIBentrySTDinterwordspacing}{\spaceskip=0pt\relax}
\providecommand{\BIBentryALTinterwordstretchfactor}{4}
\providecommand{\BIBentryALTinterwordspacing}{\spaceskip=\fontdimen2\font plus
\BIBentryALTinterwordstretchfactor\fontdimen3\font minus
  \fontdimen4\font\relax}
\providecommand{\BIBforeignlanguage}[2]{{%
\expandafter\ifx\csname l@#1\endcsname\relax
\typeout{** WARNING: IEEEtran.bst: No hyphenation pattern has been}%
\typeout{** loaded for the language `#1'. Using the pattern for}%
\typeout{** the default language instead.}%
\else
\language=\csname l@#1\endcsname
\fi
#2}}
\providecommand{\BIBdecl}{\relax}
\BIBdecl

\bibitem{B:PSE:CA:1959:Kron}
G.~Kron, \emph{Tensors for Circuits}, 2nd~ed.\hskip 1em plus 0.5em minus
  0.4em\relax New York City, NY, USA: Dover, 1959.

\bibitem{J:PSE:CA:2013:Doerfler}
F.~D{\"o}rfler and F.~Bullo, ``{Kron} reduction of graphs with applications to
  electrical networks,'' \emph{IEEE Trans. Circuits Syst. I: Reg. Papers},
  vol.~60, no.~1, pp. 150--163, Jan. 2013.

\bibitem{J:PSE:CA:2018:Kettner:1}
A.~M. Kettner and M.~Paolone, ``On the properties of the power systems nodal
  admittance matrix,'' \emph{IEEE Trans. Power Syst.}, vol.~33, no.~1, pp.
  1130--1131, Jan. 2018.

\bibitem{J:PSE:CA:2018:Kettner:2}
------, ``On the properties of the compound nodal admittance matrix of
  polyphase power systems,'' \emph{IEEE Trans. Power Syst.}, {Accepted} for
  publication. {DOI}:~10.1109/TPWRS.2018.2863671.

\bibitem{J:PSE:CA:1918:Fortescue}
C.~L. Fortescue, ``Method of symmetrical coordinates applied to the solution of
  polyphase networks,'' \emph{Trans. AIEE}, vol.~37, no.~2, pp. 1027--1140,
  Jun. 1918.

\bibitem{J:PSE:PFS:1974:Stott}
B.~Stott, ``Review of load-flow calculation methods,'' \emph{Proc. IEEE},
  vol.~62, no.~7, pp. 916--929, Jul. 1974.

\bibitem{J:PSE:PFS:1970:Meisel}
J.~Meisel and R.~D. Barnard, ``Application of fixed-point techniques to
  load-flow studies,'' \emph{IEEE Trans. Power App. Syst.}, vol.~89, no.~1, pp.
  136--140, Jan. 1970.

\bibitem{J:PSE:PFS:1968:Wallach}
Y.~Wallach, ``Gradient methods for load-flow problems,'' \emph{IEEE Trans.
  Power App. Syst.}, no.~5, pp. 1314--1318, May 1968.

\bibitem{J:PSE:CA:1968:Laughton}
M.~A. Laughton, ``Analysis of unbalanced polyphase networks by the method of
  phase coordinates. {Part 1}: System representation in phase frame of
  reference,'' \emph{Proc. IEE}, vol. 115, no.~8, pp. 1163--1172, Aug. 1968.

\bibitem{J:PSE:PFS:2018:Wang}
C.~Wang, A.~Bernstein, J.-Y. LeBoudec, and M.~Paolone, ``Explicit conditions on
  existence and uniqueness of load-flow solutions in distribution networks,''
  \emph{IEEE Trans. Smart Grid}, vol.~9, no.~2, pp. 953--962, Mar. 2018.

\bibitem{J:PSE:PFS:1982:Tiwari}
S.~N. Tiwari and L.~P. Singh, ``Six-phase (multiphase) power transmission
  systems: A generalized investigation of the load-flow problem,'' \emph{Elect.
  Power Syst. Res.}, vol.~5, no.~4, pp. 285--297, 1982.

\bibitem{J:PSE:PFS:1974:Wasley}
R.~G. Wasley and M.~A. Shlash, ``{Newton-Raphson} algorithm for 3-phase load
  flow,'' \emph{Proc. IEE}, vol. 121, no.~7, pp. 630--638, Jul. 1974.

\bibitem{J:PSE:PFS:2014:Kocar}
I.~Kocar, J.~Mahseredjian, U.~Karaagac, G.~Soykan, and O.~Saad, ``Multiphase
  load-flow solution for large-scale distribution systems using {MANA},''
  \emph{IEEE Trans. Power Del.}, vol.~29, no.~2, pp. 908--915, Apr. 2014.

\bibitem{J:PSE:SE:2000:Monticelli}
A.~J. Monticelli, ``Electric power system state estimation,'' \emph{Proc.
  IEEE}, vol.~88, no.~2, pp. 262--282, Feb. 2000.

\bibitem{J:PSE:SE:1970:Larson}
R.~E. Larson, W.~F. Tinney, and J.~Peschon, ``State estimation in power
  systems. {Part I}: Theory and feasibility,'' \emph{IEEE Trans. Power App.
  Syst.}, no.~3, pp. 345--352, Mar. 1970.

\bibitem{J:PSE:SE:1970:Schweppe:1}
F.~C. Schweppe and J.~Wildes, ``Power system static-state estimation. {Part I}:
  Exact model,'' \emph{IEEE Trans. Power App. Syst.}, no.~1, pp. 120--125, Jan.
  1970.

\bibitem{J:PSE:SE:1970:Schweppe:2}
F.~C. Schweppe and D.~B. Rom, ``Power system static-state estimation. {Part
  II}: Approximate model,'' \emph{IEEE Trans. Power App. Syst.}, no.~1, pp.
  125--130, Jan. 1970.

\bibitem{J:PSE:SE:1970:Debs}
A.~S. Debs and R.~E. Larson, ``A dynamic estimator for tracking the state of a
  power system,'' \emph{IEEE Trans. Power App. Syst.}, no.~7, pp. 1670--1678,
  Sep./Oct. 1970.

\bibitem{B:PSE:SE:2016:Paolone}
M.~Paolone, J.-Y. LeBoudec, S.~Sarri, and L.~Zanni, ``Static and recursive
  {PMU}-based state estimation processes for transmission and distribution
  grids,'' in \emph{Advanced Techniques for Power System Modelling, Control and
  Stability Analysis}, F.~Milano, Ed.\hskip 1em plus 0.5em minus 0.4em\relax
  Stevenage, HRT, UK: IET, 2016.

\bibitem{J:PSE:SE:2017:Kettner}
A.~M. Kettner and M.~Paolone, ``Sequential discrete {Kalman} filter for
  real-time state estimation in power distribution systems: Theory and
  implementation,'' \emph{IEEE Trans. Instrum. Meas.}, vol.~66, no.~9, pp.
  2358--2370, Sep. 2017.

\bibitem{J:PSE:VSA:2004:TF}
P.~S. Kundur \emph{et~al.}, ``Definition and classification of power system
  stability,'' \emph{IEEE Trans. Power Syst.}, vol.~19, no.~3, pp. 1387--1401,
  May 2004.

\bibitem{J:PSE:VSA:1992:Ajjarapu}
V.~Ajjarapu and C.~Christy, ``The continuation power flow: A tool for
  steady-state voltage stability analysis,'' \emph{IEEE Trans. Power Syst.},
  vol.~7, no.~1, pp. 416--423, Feb. 1992.

\bibitem{J:PSE:VSA:1993:Canizares}
C.~A. Ca{\~n}izares and F.~L. Alvarado, ``Point-of-collapse and continuation
  methods for large ac/dc systems,'' \emph{IEEE Trans. Power Syst.}, vol.~8,
  no.~1, pp. 1--8, Feb. 1993.

\bibitem{J:PSE:VSA:1995:Chiang}
H.-D. Chiang, A.~J. Flueck, K.~S. Shah, and N.~J. Balu, ``{CPFLOW}: A practical
  tool for tracing power system steady-state stationary behavior due to load
  and generation variations,'' \emph{IEEE Trans. Power Syst.}, vol.~10, no.~2,
  pp. 623--634, May 1995.

\bibitem{J:PSE:VSA:1997:Irisarri}
G.~D. Irisarri, X.~Wang, J.~Tong, and S.~Mokhtari, ``Maximum loadability of
  power systems using interior-point nonlinear optimization method,''
  \emph{IEEE Trans. Power Syst.}, vol.~12, no.~1, pp. 162--172, Feb. 1997.

\bibitem{J:PSE:VSA:1993:Loef}
P.-A. L{\"o}f, G.~Andersson, and D.~J. Hill, ``Voltage-stability indices for
  stressed power systems,'' \emph{IEEE Trans. Power Syst.}, vol.~8, no.~1, pp.
  326--335, Feb. 1993.

\bibitem{J:PSE:VSA:1992:Gao}
B.~Gao, G.~K. Morrison, and P.~S. Kundur, ``Voltage-stability evaluation using
  modal analysis,'' \emph{IEEE Trans. Power Syst.}, vol.~7, no.~4, pp.
  1529--1542, Nov. 1992.

\bibitem{J:PSE:VSA:1998:Prada}
R.~Prada and L.~Souza, ``Voltage stability and thermal limit: Constraints on
  the maximum loading of electrical energy distribution feeders,'' \emph{IEE
  Proc.--Gener. Transm. Distrib.}, vol. 145, no.~5, pp. 573--577, 1998.

\bibitem{J:PSE:VSA:2014:Sheng}
H.~Sheng and H.-D. Chiang, ``{CDFLOW}: A practical tool for tracing stationary
  behaviors of general distribution networks,'' \emph{IEEE Trans. Power Syst.},
  vol.~29, no.~3, pp. 1365--1371, May 2014.

\bibitem{J:PSE:VSA:2018:Kettner}
A.~M. Kettner and M.~Paolone, ``A generalized index for static voltage
  stability of unbalanced polyphase power systems including {Th{\'e}venin}
  equivalents and polynomial models,'' \emph{IEEE Trans. Power Syst.}, under
  review, available on \url{https://arxiv.org/abs/1809.09922}.

\bibitem{J:PSE:VSA:2005:Zhang}
X.-P. Zhang, P.~Ju, and E.~Handschin, ``Continuation three-phase power flow: A
  tool for voltage stability analysis of unbalanced three-phase power
  systems,'' \emph{IEEE Trans. Power Syst.}, vol.~20, no.~3, pp. 1320--1329,
  Aug. 2005.

\bibitem{PhD:PSE:SE:2017:Zanni}
L.~Zanni, ``Power-system state estimation based on {PMU}s: Static and dynamic
  approaches, from theory to real implementation,'' Ph.D. dissertation,
  {\'E}cole Polytechnique F{\'e}d{\'e}rale de Lausanne, VD, CH, 2017.

\bibitem{R:PSE:PFS:2004:IEEE}
M.~L. Baughman \emph{et~al.}, ``{IEEE} 34-node test feeder,'' IEEE PES, Tech.
  Rep., 2004.

\bibitem{J:PSE:PFS:1988:Price}
W.~W. Price \emph{et~al.}, ``Load modeling for power-flow and
  transient-stability computer studies,'' \emph{IEEE Trans. Power Syst.},
  vol.~3, no.~1, pp. 180--187, Feb. 1988.

\bibitem{J:PSE:SE:2015:Pignati}
M.~Pignati \emph{et~al.}, ``Real-time state estimation of the {EPFL}-campus
  medium-voltage grid by using {PMU}s,'' in \emph{Proc. IEEE PES Innovative
  Smart Grid Techn. Conf. (ISGT), Washington, DC, USA}, 2015, pp. 1--5.

\bibitem{Std:IEC:61869-2}
{IEC~61869-2:2012}, ``Instrument transformers, part 2: Additional requirements
  for current transformers,'' International Electrotechnical Commission,
  Geneva, GE, CH, Standard, 2012.

\bibitem{Std:IEC:61869-3}
{IEC~61869-3:2011}, ``Instrument transformers, part 3: Additional requirements
  for inductive voltage transformers,'' International Electrotechnical
  Commission, Geneva, GE, CH, Standard, 2011.

\end{thebibliography}

\end{document}